# Simulation of Social Acceptance and Information Cascades in Aquifer Thermal Energy Storage (ATES) Adoption:
## A Multiplex Agent-Based Modeling Approach

Yasunobu Wakashiro[*+]

## Abstract

Aquifer Thermal Energy Storage (ATES) offers a powerful technological option for decarbonizing urban heating and cooling systems, yet its widespread adoption is frequently hindered by strong public opposition driven by the epistemological invisibility of subsurface activities. This study develops a novel multiplex agent-based model (ABM) that couples physical heat transport (a two-dimensional advection-dispersion equation) with social network dynamics (Bayesian belief updating on a Barabási–Albert scale-free network) to elucidate the emergence mechanisms of illusory risk cascades and the dynamics of acceptance recovery through policy packages. Through 525,000 ensemble simulation runs, we quantitatively evaluate the isolated and combined effects of information provision, participatory monitoring, clear exit rules, digital twins, and economic incentives. Our analysis reveals that isolated policy interventions are largely insufficient to overcome systemic public resistance. Crucially, combining participatory monitoring with economic incentives is demonstrated to induce a non-linear 'monitoring paradox', where Bayesian updating upon anomaly detection rapidly escalates distrust, thereby nullifying the effects of financial incentives. In contrast, deploying a fully integrated, multi-layered policy package encompassing all five measures drives a macroscopic phase transition where the adoption rate jumps discontinuously at a critical threshold ($\alpha > 0.62$), successfully restoring social acceptance even under extremely high initial distrust. By mapping these mathematical model dynamics onto frameworks of 'procedural justice' and 'distributive justice', this work provides a theoretical and practical roadmap for governing invisible subsurface renewable energy infrastructures.

## 1. Introduction

### 1.1. Urban Decarbonization and Aquifer Thermal Energy Storage Technology

It has been reported that the impacts of climate change on urban areas are severe not only in terms of physical and economic risks, but also regarding their adverse effects on economic activities and human life [1]. As summer heat island phenomena and global warming explosively increase urban cooling demand [2], while space conditioning demands further emit greenhouse gases [3], city-led decarbonization has become essential [4]. As a strategic path toward urban decarbonization, thermal decarbonization—particularly thermal energy storage—has drawn significant attention in Europe; studies have highlighted that combining district heating with thermal energy storage offers the most economical solution [5], and that fourth-generation district heating systems utilizing low-temperature waste heat and geothermal energy provide exceptional efficiency among thermal storage technologies [6].

* Faculty of Economics, Setsunan University
+ E-mail: yasunobu.wakashiro@setsunan.ac.jp

Aquifer Thermal Energy Storage (ATES) is a technology that stores and utilizes thermal energy in groundwater located within deep permeable geological formations. This technology regulates seasonal thermal demand by accumulating summer heat in groundwater for winter heating and utilizing winter coldness for summer cooling. It has been shown to be well-suited for large-scale seasonal thermal storage from the perspectives of energy efficiency and groundwater conservation, while dramatically enhancing overall urban energy efficiency when integrated into district heating and cooling networks [7]. ATES implementation is actively progressing in countries such as the Netherlands and China [8], and global implementation records and technical potentials are being actively researched to support future technology penetration [9] [1].

## 1.2. Social Acceptance as a Barrier to Renewable Energy Investment

[13] reviewed 174 academic papers on ATES published between 2020 and 2025, confirming that the majority of studies address economic feasibility. While energy investments, including ATES, cannot proceed without economic viability, it has been highlighted that even when technical and economic challenges are resolved, "non-technical barriers (social and institutional)," particularly social acceptance, frequently impede development [14], [15], [16], [17]. In case studies of renewable energy investment, [18] demonstrated that the primary barriers to renewable energy diffusion in the United States were not costs, but rather "sociotechnical lock-in" and "resistance from vested interests." Similarly, through a policy review in Australia, [19] noted that even with financial support, distrust in policy consistency (non-economic risk) dampens investment. Regarding wind power, [20] pointed out that non-economic barriers, such as a lack of social recognition, are critical to investment and development, while [21] showed that implementation is determined not by cost, but by the "dynamics of planning systems and opposition movements."

[1] Examples include [10], which evaluated optimization methods for large-scale ATES in urban areas and its potential in densely populated areas, as well as [11], [12], which analyzed the impact of groundwater flow on ATES efficiency and proposed optimal layout designs.

Unlike visible surface energy infrastructures—such as wind turbines, mega-solar farms, or high-voltage transmission lines—ATES is developed and operated deep underground, making direct observation impossible for nearby residents. Furthermore, there is a theoretical possibility that nearby ATES operations could induce bacterial growth in groundwater due to heat transport, or land subsidence caused by groundwater pressure changes. Although these occurrences are extremely improbable under proper management and operation, verifying this directly within realistic cost constraints is difficult. Consequently, when subtle noise—such as natural groundwater temperature fluctuations or minor hydraulic head differences—is detected, it can evoke associations with catastrophic disasters (e.g., land subsidence, groundwater contamination, or induced seismicity), even in the complete absence of actual physical harm.

## 1.3. Information Transmission and Decision-Making

Extensive research has analyzed how risk perceptions of events or technologies, amplified through information transmission channels, can trigger public panic or policy stalls, and how such information irrationally distorts individual decision-making [22], [23], [24]. Social movements and opposition behaviors are characterized as "complex contagions," which amplify within network clusters (in modern contexts, social media echo chambers). Scholars have emphasized that fear-inducing misinformation (fake news) spreads faster, deeper, and more broadly on social media than scientific truths, and that the principle of "group polarization" drives like-minded online groups toward more extreme actions (radical opposition movements). Furthermore, studies have demonstrated how distrust of scientific narratives becomes entrenched as a solid "collective identity" within conspiracy-minded communities [25], [26], [27], [28], [29]. When minor noise observed after ATES deployment amplifies rumors of impending catastrophic disaster and fuels anxiety, residents' distrust and risk perceptions become governed not by objective, physical states, but by subjective risk constructions, cognitive biases, and unverified social rumors.

Real-world instances where opposition movements against energy deployment and operation arose via the internet and social media include smart meter rollouts, where rumors of "health hazards" and "privacy violations" diffused across social networks and substantially delayed implementation [30], as well as shale gas development, where online amplification of fear and conspiracy theories shaped widespread public anxiety [31]. In wind energy cases, reported health complaints were triggered not by physical harm from the generation facilities themselves, but by the "propagation of fear via rumors and media (the nocebo effect)," ultimately derailing project plans [32]. In a review of barriers to renewable energy adoption, [33] identified the absence of information governance and social acceptance as the paramount barrier in the social media era.

Entities undertaking energy investments perceive "regulatory ambiguity" and "social resistance" as primary investment risks [34], with delays in administrative permitting processes—often stemming from local community opposition—constituting severe regulatory uncertainty [35]. Empirical evidence shows that non-financial risks unique to renewable energy projects ultimately force project cancellations and sunk costs [36], and that investors withhold capital in high social risk environments characterized by opposition-driven derailments, even when financial subsidies are present [37]. From a valuation methodology perspective, the uncontrollability of non-market risks such as social opposition repels investors [38]; real options theory further demonstrates that under high volatility in public distrust, investors rationally adopt a "wait and see" strategy to defer investment [39]. Taking these factors into account, designing institutional frameworks that mitigate investment risks—including social and policy frictions—is paramount to promoting renewable energy deployment [40].

Studies utilizing ABM have formalized how Not In My Back Yard (NIMBY) movements originate from localized individual dissatisfaction and escalate into macroscopic opposition campaigns through "network dynamics" on social media platforms [41], [42]. As a consequence of such escalated opposition, non-financial risks such as public protests and litigation drive up the cost of capital [43], social friction and policy volatility elevate hurdle rates to render projects unprofitable, and "schedule delay risks" exert a fatal impact on capital recovery [44]. Crucially, public distrust and opposition constitute a fatal risk to financing not only for private enterprises, but also for municipal and community-led projects [45].

Regarding policy interventions to reduce social risk, the traditional "information deficit model"—which assumes that disproving groundless rumors through timely disclosure of correct information will dissipate concern—has been refuted by [46], who demonstrated that opposition persists unless underlying deficits in institutional trust are addressed. This is further supported theoretically by [47], who illuminated how "alternative narratives" (rumors) contradicting official announcements are systematically produced and consumed, particularly during crisis situations. Underpinning this phenomenon is the recognition that public distrust of science and technology stems not from fear of the technology itself, but from distrust of the governing authorities managing it—a sentiment readily amplified on social media [48]. While recent scholarship advocates for participatory governance or economic compensation (benefit sharing), the vast majority evaluates policies in isolation. In complex sociotechnical systems, however, policy interventions interact non-linearly. For instance, combining participatory monitoring with financial compensation can generate unexpected trade-offs, such as a "monitoring paradox" where heightened surveillance amplifies public anxiety rather than trust, thereby nullifying compensation effects.

### 1.4. Purpose of This Study and Structure of the Article

To address these critical gaps, this study constructs a novel multiplex agent-based model (ABM) coupling a two-dimensional physical aquifer heat advection-dispersion model (physical layer) with a complex social network model driven by Bayesian belief updating (social layer). Operating as an "exploratory ABM" to uncover non-linear interactions and phase dynamics between physical transport and multi-layered policy packages, this study addresses three core research questions:

- How do subtle subsurface physical noises trigger illusory risk cascades across social networks?
- To what extent do interventions designed to mitigate cascades succeed, and what consequences emerge from deploying multiple policy interventions simultaneously?
- Can a multi-layered policy package successfully restore social acceptance under conditions of extremely severe initial public distrust?

## 2. Methodology

We formulate the urban ATES adoption ecosystem as a multiplex system comprising two interconnected layers: (1) a spatial physical grid layer representing the subsurface thermal aquifer environment, and (2) a topological social network layer representing the communication structure among local residents. The physical layer governs the spatiotemporal evolution of temperature plumes and hydraulic pressure changes, while the social layer governs the propagation of risk perceptions, confirmation biases, and decision-making dynamics regarding ATES system adoption or opposition movements. As pointed out by [49], "timescale asymmetry" generates critical risks in coupled human-environment systems; namely, while natural phenomena evolve slowly, human societies react precipitously to perceived hazards [50]. In sharp contrast to thermal advection and dispersion in aquifers—where thermal plumes require months to years to migrate tens of meters—social information diffusion via modern digital networks and local community channels occurs on rapid timescales of hours to days. When operational noise occurs, cascades of distrust and rumors diffuse across the social network orders of magnitude faster than underlying physical thermal movement can stabilize or dissipate. Existing energy acceptance models have either treated social acceptance as a static parameter or overemphasized physical simulations lacking dynamic, real-time information feedback loops.

Furthermore, it has been highlighted that the inherent invisibility of the subsurface amplifies public anxiety, triggering cascades of social unrest. [51] demonstrated in the context of hydraulic fracturing (subsurface development) that "subsurface invisibility" itself instills distinct epistemological fear and conspiracy theories among residents. As emphasized by [26], unlike simple viral contagion, rumors and participation in social opposition movements represent "complex contagions," propagating through peer pressure from multiple neighbors and suddenly spreading across the entire network at a critical threshold (phase transitions or avalanches). When combined with epistemological anxiety over invisibility, this characteristic induces psychosomatic panic (the nocebo effect) driven by rumor propagation—such as claims that "people will fall ill"—even in the complete absence of physical harm [32].

The multiplex ABM combining physical and social layers in this study adopts an approach that models natural environmental dynamics and human social network decision-making as distinct yet coupled layers. Previously proposed models include multiplex frameworks linking "ecological networks (physical layer)" sharing common resources with "social networks (social layer)" where agents exchange information [52], models coupling ATES thermal interference (subsurface physical constraints) with operator self-organization (social and economic behaviors) [53], and ABMs integrating physical environments with human irrational behavioral biases during energy system transitions [54]. Drawing upon [54] and [52], our model mathematically integrates these dimensions within a multi-layered network to quantitatively evaluate the impacts of policy interventions, such as participatory governance, on phase transitions.

Furthermore, building on studies evaluating the comparative effectiveness of multiple policy measures [55], analyzing how the presence or absence of participatory governance affects the success or failure (social acceptance) of spatial energy planning [56], examining how ownership structures alter local resistance risks and bounce back onto financing costs [57], and demonstrating that local "ownership and benefit-sharing" are indispensable for evading social resistance risks caused by external capital monopolization [58], this study conducts ABM-based simulations to evaluate whether policies such as "participatory monitoring committees" and "benefit sharing" function as viable solutions and to further explore comprehensive policy packages.

## 2.1. Basic Model Architecture

The present model is a multiplex network consisting of a physical layer ($\tau_{physical}$) that computes heat transport on a two-dimensional spatial grid, and a social layer ($\tau_{social}$) through which information propagates across a scale-free network [59], [60]. Multiplex networks, which integrate and analyze different types of interactions, represent a standard multilayer approach in complex network science [60]. In an approach superimposing a "two-dimensional grid" and a "social network," it is possible to physically model dynamics unique to multiplex networks that cannot be observed in a single network, thereby demonstrating how slow changes in the physical layer trigger avalanches in the social layer [59].

Regarding the social layer $\tau_{social}$ representing rumor propagation on social networking services (SNS), the structure of SNS is approximated as a scale-free network, and rumor propagation across this network is modeled accordingly. The network architecture adopts a Barabási–Albert (BA) scale-free topology with $N = 10{,}000$ agents featuring a small number of massive hubs (influencers) following real-world social network and web structures, with an average degree of new node attachments of $m = 5$ (to mimic the formation processes of peer pressure and echo chambers characteristic of modern SNS) [61]. This network reproduces the heavy-tailed degree distribution of modern digital information networks. By solving the propagation equation across this network, the model reproduces behavior in which rumors spread across the entire system within an extremely short timeframe due to the presence of hubs in scale-free networks [62].

Agents in this model do not react solely as a function of physical thermal interference (actual harm); rather, they perform Bayesian inference grounded in subjective probability. Each agent $i$ maintains a subjective distrust belief $P_i(H_1)$ regarding ATES safety. Upon detecting a noisy anomaly or rumor signal $E$, the agent updates distrust beliefs according to a Bayesian updating rule incorporating a confirmation bias coefficient $\gamma$. Notably, agents in regions that have experienced past environmental traumas (e.g., groundwater contamination or land subsidence) maintain extremely high initial distrust due to environmental stigma [63]—that is, a high prior probability $P(H_0)$ [64]:

$$P_i(H_1|E) = \frac{P(E|H_1)^{\gamma} \cdot P_i(H_1)}{P(E|H_1)^{\gamma} \cdot P_i(H_1) + P(E|H_0)^{\gamma} \cdot \left(1 - P_i(H_1)\right)}$$

The physical layer $\tau_{\text{physical}}$ representing ATES thermal transport is modeled using the advection-dispersion equation (ADE). Based on the formulation by [65], which defined partial differential equations for fluid and heat movement via advection (transport by groundwater flow) and dispersion (diffusion) in porous media such as subsurface aquifers, this physical advection-dispersion model simulates how thermal plumes discharged from actual ATES systems disperse along groundwater flow and reach other users' wells [66]. Thermal energy transport in a confined aquifer is described by the two-dimensional heat advection-dispersion equation in porous media:

$$\frac{\partial T}{\partial t} = D_x\left(\frac{\partial^2 T}{\partial x^2}\right) + D_y\left(\frac{\partial^2 T}{\partial y^2}\right) - v_x\left(\frac{\partial T}{\partial x}\right) - v_y\left(\frac{\partial T}{\partial y}\right) + S_{\text{thermal}}$$

where $T(x, y, t)$ denotes the spatial temperature field, $D_x$ and $D_y$ are effective thermal dispersion coefficients, $v_x$ and $v_y$ are groundwater seepage velocity vectors, and $S_{\text{thermal}}$ represents source/sink terms corresponding to heat injection and extraction wells. The physical domain is discretized into a $100 \times 100$ spatial finite-difference grid.

## 2.2. Illusory Avalanches and Bayesian Updating

When confirmation bias [67] is induced within human cognitive processes, agents misattribute subtle temperature fluctuations and routine environmental noise unrelated to ATES as "damage caused by ATES," self-amplifying their distrust through Bayesian updating grounded in illusory correlation [68]. In particular, the extremely high initial distrust of agents in regions that have experienced past environmental traumas can trigger powerful confirmation bias. When self-amplified distrust surpasses the threshold for opposition behavior, agents transition into a BANANA state (Build Absolutely Nothing Anywhere Near Anything; [69]) and transmit "rumors of damage" across the social network. As threshold models demonstrate [70], these rumors cascade by lowering the behavioral thresholds of neighboring agents. Consequently, a rejection response spreads throughout the entire system despite the complete absence of actual physical harm. Local concerns are thereby radicalized through information sharing across the network, further broadcasting "rumors of damage" across social communication channels. Individual behavior is governed not by absolute damage amounts, but rather by "how many surrounding peers are engaging in opposition," which lowers behavioral thresholds in a chain reaction; as [71] showed, the behavior of a small number of agents on the network (the initial misattributors), once surpassing a critical point, triggers a sudden system-wide cascade (avalanche). The rumor thereby sequentially lowers the behavioral thresholds of neighboring agents through complex contagion [26]. This gives rise to an "illusory avalanche," wherein the entire system becomes engulfed in an opposition movement even though physical harm never reaches the network.

Cascades of opposition movements triggered by nocebo effects and related phenomena in wind and solar power generation ignite by relying on clear "visual anchors." Residents use visible, large-scale infrastructure as a focal point to post-hoc associate their ill health or landscape changes [32]. In such cases, the cascade's reach to distant residents who cannot visually verify the infrastructure tends to be spatially constrained. In contrast, ATES, which utilizes subsurface space, lacks visual anchors [51]. Because ATES cascades are unconstrained by visual evidence, they proceed relying solely on imagination and prior biases. Driven by a fundamental fear that "something is occurring beneath our feet out of sight," residents connect unrelated routine noise (such as minor temperature fluctuations or well water turbidity) to the system while disregarding spatial distance. In our model, this "absence of visual constraints" is embedded within the Bayesian updating algorithm as a mechanism that permits rumor propagation (cascades) across broader networks at significantly lower thresholds than visible infrastructure.

## 2.3. Agent Decision Rules and Irreversible Opposition (BANANA State)

Agents evaluate utility across three operational states: Adoption, Neutral, and Protest. The net protest utility $U_{\text{protest},i}$ is formulated as follows:

$$U_{\text{protest},i} = w_{\text{distrust}} \cdot P_i(H_1|E) + w_{\text{social}} \cdot \sum S_j - U_{\text{cost}} + U_{\text{penalty}}$$

Grounded in Slovic's (1993) principle of trust asymmetry ("trust is hard to gain and easy to lose"), if cumulative protest utility exceeds a critical threshold ($U_{\text{protest}} > 15$), the agent transitions into a permanent opposition state—termed the BANANA state (Build Absolutely Nothing Anywhere Near Anything)—from which state recovery is precluded.

## 2.4. Operationalization of Policy Interventions (Policies A–E)

Interventions designed to build institutional trust are defined as mathematical parameter adjustments (intervention strength $\alpha \in [0,1]$) within the agent decision-making algorithm (behavioral rules). The theoretical rationale and mathematical implementation of each policy are as follows. Policy package scenarios (e.g., information-driven, governance-driven, benefit-sharing-driven, and full package) combining these measures are established, and Monte Carlo simulations are executed treating environmental trauma levels and policy intervention strength as variables.

### 2.4.1 Preemptive Communication to Reduce Prior Distrust (Policy A)

From the perspective of cognitive psychology, while correcting misinformation after it has taken root is notoriously difficult, debiasing through preemptive explanation of misinformation mechanisms effectively mitigates initial human biases [72].

### 2.4.2 Elimination of "False Positives" via Participatory Monitoring Committees (Policy B)

While technical data produced exclusively by experts can evoke institutional distrust, involving residents from the planning stage enhances information credibility [73]. Integrating local community knowledge with expert data has been shown to prevent unproductive conflicts and rumor generation, particularly in local environmental issues [74]. Direct resident participation in data monitoring reduces the likelihood $l_{\text{false}}$ of false positive misidentifications where information noise is perceived as actual damage [75]. However, if an operational anomaly signal is detected under near-zero false positive conditions, distrust rapidly spikes to 1.0 (the monitoring paradox).

### 2.4.3 Avoiding Worst-Case Scenarios via Clear Exit Rules (Policy C)

In the governance of common-pool resources, prior consensus on exit rules to prevent catastrophic errors is indispensable [76]. Institutional guarantees of reversibility establish an upper bound on the worst-case expected loss (fear) associated with unfamiliar technologies. Codifying exit rules mitigates the weight of the harm penalty $U_{\text{penalty}}$` in the utility function, thereby suppressing extreme risk-averse behaviors among agents.

### 2.4.4 Real-Time Data Disclosure and the Information Transparency Paradox (Policy D)

Disclosing real-time data to enhance transparency is expected to suppress factually ungrounded rumors. Conversely, scholars have cautioned that excessive monitoring can induce hypersensitivity to minor operational noise [77]. While mitigating rumor impact, it simultaneously heightens agent sensitivity to immediate thermal fluctuations.

### 2.4.5 Economic Benefit Sharing with Local Residents (Policy E)

In renewable energy projects, local distribution of economic benefits fosters resident ownership, dramatically boosting acceptance [78]. Economically, this value sharing represents not a mere bribe, but a structural enhancement of institutional acceptability (utility). Direct benefits, such as free thermal energy or financial dividends, function as strong incentives outperforming social friction and opposition behavior [79].

### 2.4.6 Simulation Protocol, Parallel Architecture, and Policy Operationalization

To derive the expected values of macroscopic phase transition phenomena, a large-scale, high-resolution Monte Carlo ensemble simulation was conducted. To eliminate topological noise arising from the spatial allocation of initial adopters and the generation of scale-free network hub structures, 30 independent runs with distinct random seeds were executed for each parameter set, computing the ensemble average of system adoption rates. The simulation horizon was set to $t_{\max} = 400$ timesteps to confirm complete system convergence. To precisely map the effects of policy interventions, the parameter space of initial prior distrust $P(H_0) \in [0.0,1.0]$ and policy intervention strength $\alpha \in [0.0,1.0]$ was discretized into 50 intervals each, varying across a $50 \times 50$ matrix. Evaluating this across seven policy scenarios yielded a grand total of 525,000 full simulation runs, including ensemble trials. To manage the immense computational complexity, spatial calculations for the advection-dispersion equation in the physical layer were fully vectorized, and parallel distributed processing across 32 logical threads was implemented under an architecture minimizing inter-process communication (IPC) overhead. This established a rigorous foundation for seamlessly evaluating dynamics from microscopic cognitive biases to macroscopic social phase transitions at high spatial and statistical resolution.

The simulation sequence proceeds as follows. Computation first advances in the physical layer. Thermal energy generated beneath agents adopting ATES migrates along groundwater flow, simulated via the two-dimensional finite-difference advection-dispersion equation (ADE) parameterized with an advection velocity coefficient of 0.3 (thermal transport via groundwater flow) and a dispersion coefficient of 0.05 (thermal diffusion into surrounding media). Upon completing physical layer calculations, computation transitions to the social layer (agent decision-making). Each agent observes the system state and acts according to policy implementations as follows. Upon detecting either "actual thermal energy beneath their feet" or "rumors from surrounding opposition peers," agents update their distrust beliefs using

Bayes' theorem and react across the social network. Agents compare the "utility of adoption" against the "utility of protest." If net protest utility exceeds the critical threshold, the agent transitions into a permanent opposition state (BANANA: state -1), from which state recovery or reversal is precluded.

The operational impacts of each policy intervention on agent behavioral parameters are specified as follows:

- Policy A (Preemptive Communication): Reduces the initial prior distrust probability $P(H_0)$ by up to 50%.
- Policy B (Participatory Monitoring Committee): Reduces the false positive rate ($l_{\mathrm{false}}$) by up to 80%.
- Policy C (Clear Exit Rules): Reduces the expected loss penalty by up to 80%.
- Policy D (Digital Twin Dashboard): Reduces the weighting of rumor influence by up to 80%, while simultaneously lowering the detection threshold for physical noise by up to 50%.
- Policy E (Economic Benefit Sharing): Adds up to +20 to expected adoption utility and subtracts up to -10 from net protest utility.

# 3. Results

By executing Monte Carlo ensemble simulations, we generated 15 phase diagram matrices with the initial prior distrust probability $P(H_0)$ and policy intervention strength α as variables (Figure 3.1). Comparing individual policy measures revealed that isolated policies have clear structural limitations, and that simultaneous implementation of multiple policy measures induces non-linear interactions.

## 3.1. Efficacy of Isolated Policy Interventions and the Linear Frontier of Economic Incentives

We first describe the effects of implementing Policies A through D in isolation. When Preemptive Communication (Policy A) is deployed alone, adoption can be anticipated even under high trauma levels; however, the adoption rate itself remains at the same level as the baseline (no policy) scenario. Participatory Monitoring (Policy B) yields adoption rates comparable to the baseline at low intervention strengths, but adoption declines as intervention strength increases. Codifying Clear Exit Rules (Policy C) produces adoption rates identical to the baseline scenario. The Digital Twin Dashboard (Policy D) achieves adoption under high trauma levels only when policy strength is exceptionally high; at lower policy strengths, however, the adoption rate falls below the baseline.

In contrast, Economic Benefit Sharing (Policy E) forms a diagonal 'adoption band' (a bright yellow frontier) extending upward in proportion to policy strength α. This indicates that because Policy E directly adds a positive value to the utility function, adoption succeeds linearly only at specific equilibrium points where the level of prior distrust (negative term) and the compensation amount (positive term) balance each other out.

Overall, single defensive policies (Policies A–D in isolation) fail to achieve social acceptance in regions where initial distrust exceeds baseline levels, leaving adoption rates below 1% ($R_{\text{adopt}} < 0.01$). Even Policy E—the sole policy capable of driving adoption in isolation—has its efficacy strictly bounded by a diagonal linear frontier; in high-risk regimes where $P(H_0) > 0.40$, presenting even a +20 financial incentive is overwhelmed by cumulative protest utility.

## 3.2. Combined Defensive Interventions Without Economic Incentives (Policies A–D)

Next, we examine the effects of combining defensive policies A through D in the absence of Economic Benefit Sharing (Policy E). When Preemptive Communication (Policy A) was combined with the Digital Twin Dashboard (Policy D), the low adoption rate observed at low policy strength under Policy A in isolation disappeared, while adoption was significantly enhanced under extremely high Digital Twin policy strength. When Participatory Monitoring (Policy B) was combined with Clear Exit Rules (Policy C), the adoption observed at high policy strength under Policy C in isolation completely vanished. Combining all defensive policies (A–D) yielded results nearly identical to the combination of Policy A and Policy D.

The primary objective of these defensive policies is either to proactively reduce initial resident distrust (preemptive communication, clear exit rules) or to remove drivers of opposition by disclosing operational states (participatory monitoring, digital twin dashboards). While these measures eliminate negative perceptions (demerits), they fail to generate positive adoption benefits (merits). Consequently, defensive combinations alone remain incapable of pushing agent utility beyond the decision-making threshold for adoption.

## 3.3. Pairwise Policy Interactions with Economic Benefit Sharing (Policy E)

### 3.3.1 Simultaneous Implementation with Participatory Monitoring (Policy B) or Clear Exit Rules (Policy C): The Monitoring Paradox

When Participatory Monitoring (Policy B) or Clear Exit Rules (Policy C) was combined with Economic Benefit Sharing (Policy E), the distinct adoption band observed under Policy E in isolation completely vanished. This indicates that participatory monitoring can inadvertently nullify the positive incentives of benefit sharing. Recalling that participatory monitoring is designed to reduce the probability of misidentifying noise as harm ($l_{\text{false}}$), Bayesian inference mathematics dictates that when an anomaly (even if mere noise or rumor) is detected under conditions where the false positive rate is extremely low (where the denominator approaches zero), the posterior distrust probability spikes sharply to 1.0, generating absolute conviction among agents. In other words, disclosing flawless technical data causes residents to react with extreme volatility to low-probability operational noise, which in turn severely undermines the efficacy of complementary policies.

### 3.3.2 Simultaneous Implementation with Preemptive Communication (Policy A) or Digital Twin Dashboard (Policy D)

When Economic Benefit Sharing (Policy E) was deployed simultaneously with Preemptive Communication (Policy A) and the Digital Twin Dashboard (Policy D), an extensive high-adoption region formed, demonstrating strong synergistic policy performance.

The emergence of the bright yellow band (a stable regime of high adoption) can be explained through the following mechanism: the reduction of prior bias via Policy A and the positive utility added by Policy E operate in a substitutive and complementary relationship. As policy strength α increases, these measures simultaneously deduct from the negative term (prior trauma) and add to the positive term (economic gain) in the utility function, driving adoption upward. While enhancing Policy A and Policy E alone is insufficient to achieve adoption in high-trauma regimes, introducing Policy D beyond a critical threshold ($\alpha \approx 0.60$) physically severs rumor propagation pathways across the network by exceeding the percolation threshold.

The mechanism underlying the purple and orange gradient bands (intermediate and localized adoption) represents a transition boundary between system-wide adoption and cascade collapse. Near the boundary where α falls slightly short of the yellow core regime, Digital Twin intervention strength remains insufficient, leaving partial rumor transmission pathways intact across the network. In this transition regime, adoption outcomes depend heavily on spatial agent allocation and the structural location of network hubs (stochastic variance across random seeds). In certain simulation runs, groups of agents attracted by financial incentives form localized "pro-adoption clusters" that survive; in other runs, agents are swallowed by rumor cascades and transition into the BANANA state. Because our program computes an ensemble average across 30 Monte Carlo runs, the purple and orange hues do not represent partial acceptance by all agents, but rather reflect that adoption succeeded in a subset of trials where rumors were successfully contained.

That this smooth gradient band appears exclusively in the Policy A + D + E combination can be attributed to the absence of Participatory Monitoring (Policy B) and Clear Exit Rules (Policy C). Because Policy B is absent, agents do not exhibit hyper-reactive sensitivity to operational noise that would otherwise burn off economic incentives. Instead, remaining relatively insensitive to noise, agents evaluate prior bias reduction (Policy A), rumor discounting (Policy D), and direct economic gain (Policy E). Consequently, the system avoids binary "all-or-nothing" cascade collapses, tracing adoption dynamics analogous to a continuous phase transition in physics—progressing smoothly from purple (partial adoption) to orange (moderate adoption) and yellow (full adoption) as policy strength $\alpha$ increases.

Furthermore, when Policy A and Policy E were deployed together without Policy D, an interference pattern emerged wherein the adoption band bifurcated into two distinct stripes. This non-linear resonance occurs because the direct reduction of prior trauma (Policy A) and the augmentation of utility (Policy E) experience a phase shift, causing the Bayesian updating threshold to interfere non-linearly with the decision-making threshold within specific parameter regimes.

### 3.4. Macroscopic Phase Transition Under Full Policy Package Integration (Policies A–E)

When all five policy interventions (Policies A–E) were deployed simultaneously, a sharp, vertical boundary line emerged once policy intervention strength surpassed a critical threshold. In the full-package regime where $\alpha > 0.62$, adoption rapidly spiked to near-complete levels, even under extreme initial trauma conditions ($P(H_0) \approx 0.90$).

This non-linear jump occurs because the destructive monitoring paradox observed under the B + E pair is effectively neutralized by the simultaneous deployment of Policy A and Policy D, which weakens rumor propagation power across the network. Only when cognitive bias mitigation (Preemptive Communication + Digital Twin), risk capping (Participatory Monitoring + Clear Exit Rules), and substantial utility enhancement (Economic Benefit Sharing) operate in concert beyond the critical threshold ($\alpha \approx 0.62$) is the transition of key network hub agents into the BANANA state prevented. Consequently, information cascades across the social network are physically severed, inducing a rapid, systemic shift in adoption across the entire population.

Crucially, an abrupt, discontinuous **macroscopic phase transition** occurs exclusively under the fully integrated, multi-layered policy package. The instant intervention strength crosses the critical threshold ($\alpha_{\text{critical}} > 0.62$), the system adoption rate jumps precipitously from $R_{\text{adopt}} < 0.01$ to $R_{\text{adopt}} > 0.95$, forming a robust vertical phase boundary that maintains stable social acceptance even under severe initial community distrust ($P(H_0) = 0.90$).

**Fig.1 Effects of Policy Packages on ATES Adoption**

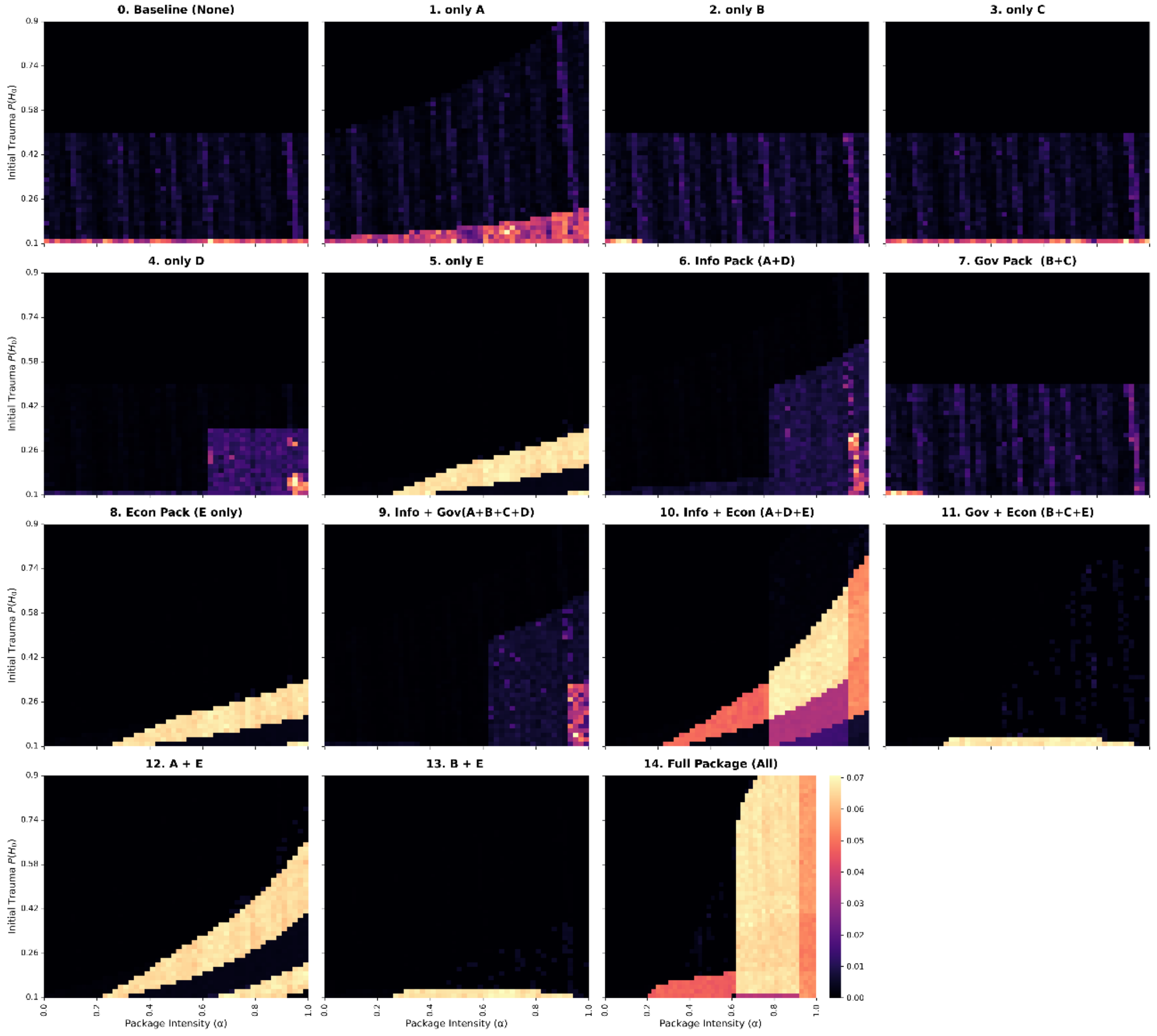

## 3.5. Detailed Pairwise Policy Interactions Across Trauma Regimes

In the preceding section, policy interference, offset, and complementary relationships were observed. To systematically analyze these pairwise interactions, we held the initial prior trauma level constant across discrete values ($P(H_0) = 0.2, 0.4, 0.6$, and $0.8$).

Notably, across all resulting phase diagram matrices, the maximum value of the colorbar remains constrained to extremely low orders of magnitude (ranging from $10^{-3}$ to $10^{-2}$). This fact strongly underscores a fundamental premise: no pairwise combination of two policy interventions, regardless of their individual strengths, is capable of triggering a macroscopic phase transition to achieve full system adoption (100% adoption rate). Below, we analyze the specific characteristic interactions observed between key policy pairs.

### 3.5.1 Preemptive Communication (Policy A) vs. Economic Benefit Sharing (Policy E)

As evidenced particularly in the panels for trauma levels $P(H_0) = 0.2$ and $0.4$, an adoption band characterized by a downward-sloping boundary is formed. This indicates that as the strength of Preemptive Communication decreases, the strength of Economic Benefit Sharing required to induce an equivalent level of adoption increases linearly. Furthermore, under extreme trauma conditions ($P(H_0) = 0.8$), minor signs of adoption emerge exclusively within a confined upper-right region of the parameter space (where both Preemptive Communication and Economic Benefit Sharing operate at high intensity).

This demonstrates that substitutability exists between the "cognitive dimension" (Bayesian prior probability) and the "utility dimension" (financial incentives) within the agent decision-making algorithm. Providing preemptive information lowers initial trauma, thereby reducing the expected value of the harm penalty and enabling even modest financial compensation to shift net utility into positive territory. Conversely, if initial trauma remains unaddressed, massive economic compensation is required to forcibly elevate net utility. Mathematically, these two policy dimensions maintain a linear complementary and substitutive relationship.

**Fig.2 Synergistic Effects: Policy A vs Policy E**

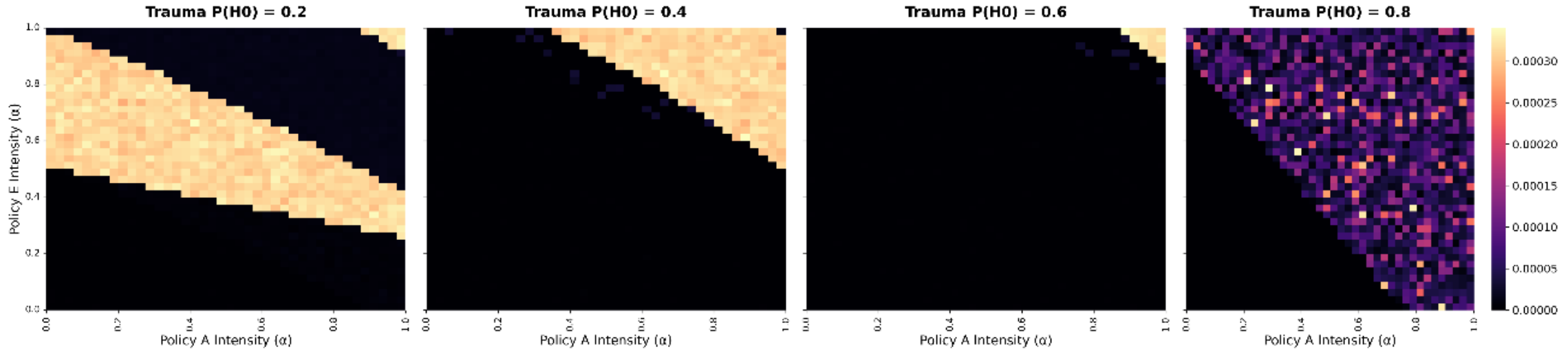

### 3.5.2 Participatory Monitoring (Policy B) vs. Economic Benefit Sharing (Policy E)

Under low trauma ($P(H_0) = 0.2$), adoption occurs exclusively within the upper-left region, forming an upward-sloping frontier. Even when Economic Benefit Sharing is maximized ($\alpha = 1.0$), elevating Participatory Monitoring strength beyond 0.5 causes adoption to abruptly drop to zero (a completely dark regime). Under high trauma conditions ($P(H_0) \geq 0.6$), the entire parameter space remains submerged in darkness.

Participatory monitoring functions to lower the probability of misidentifying unrelated noise as physical damage ($l_{\text{false}}$). However, if noise is detected when initial trauma (strong prior bias) remains while the false positive probability has been drastically reduced, the denominator in Bayes' theorem shrinks, causing the posterior distrust probability to instantly spike to 1.0 (absolute conviction). In essence, participatory monitoring under these conditions renders residents hyper-sensitive to subtle anomalies, causing the economic utility provided by benefit sharing to be instantly wiped out by a single Bayesian belief update.

**Fig.3 Synergistic Effects: Policy B vs Policy E**

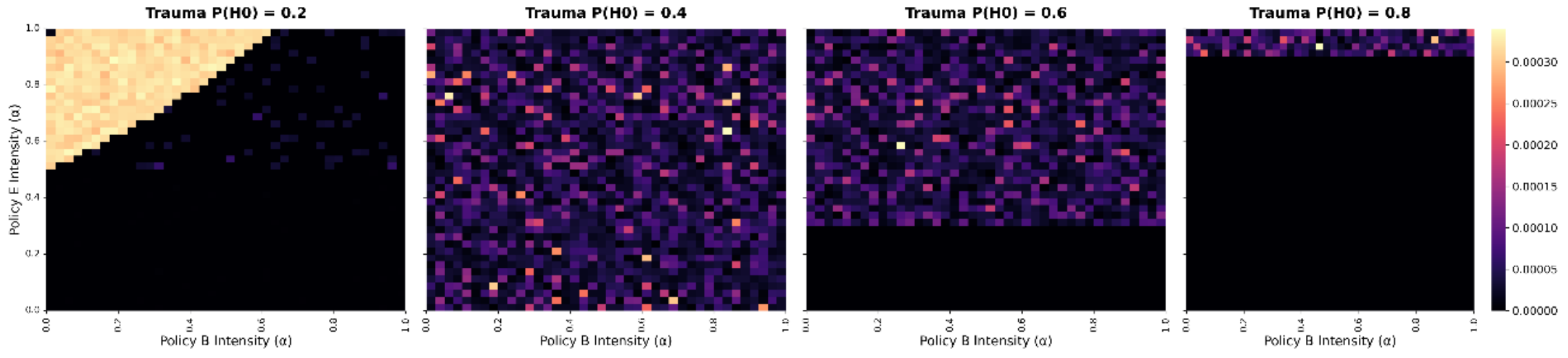

## 3.6. Synthesis and Comprehensive Policy Evaluation

- **Preemptive Communication (Policy A):** Resets cognitive biases and serves as the foundational groundwork for all complementary measures, though it generates no positive adoption utility in isolation. However, it maintains a strong linear complementary relationship with Economic Benefit Sharing (Policy E); mitigating prior trauma via preemptive communication lowers the minimum financial compensation required to achieve adoption.
- **Participatory Monitoring (Policy B):** While expected to enhance information credibility, Bayesian updating dynamics dictate that under near-perfect monitoring where false positives are minimized, detecting even minor noise causes resident distrust to instantly jump to 1.0 (absolute conviction), completely neutralizing financial incentives. Nevertheless, Participatory Monitoring functions effectively when Preemptive Communication and Digital Twins (Policy D) operate in tandem to suppress noise detection and rumor diffusion across the network.
- **Clear Exit Rules (Policy C):** Generates zero positive adoption utility in isolation; however, for high-trauma residents who perceive potential operational accidents as catastrophic ("infinite harm"), codifying exit rules is paramount by establishing a mathematical ceiling on the maximum expected loss. This institutional ceiling is what allows the positive utility of benefit sharing to outweigh negative terms and cross the decision-making threshold.
- **Digital Twin Dashboard (Policy D):** Although involving a trade-off that heightens local noise sensitivity, the instant policy strength $\alpha$ surpasses a critical threshold ($\alpha \approx 0.60$), it severs rumor transmission pathways running through scale-free network hubs. It thus serves as the ultimate firewall against cascades, preventing positive effects from Preemptive Communication and Benefit Sharing from being swallowed by "illusory avalanches" before diffusing across the network.
- **Economic Benefit Sharing (Policy E):** Directly generates positive utility, standing as the sole policy capable of promoting adoption in isolation. However, in regimes where prior trauma $P(H_0) > 0.40$, even massive financial compensation is overwhelmed by distrust and fear. To successfully drive adoption through benefit sharing, Preemptive Communication must clear cognitive obstacles, Clear Exit Rules must activate institutional safety nets, and participatory monitoring must be deployed within an integrated policy framework.

# 4. Discussion

## 4.1. Theoretical Implications of Non-Linear Policy Interactions

The emergence of the "monitoring paradox" (where hyper-transparent participatory monitoring nullifies economic benefits) and the abrupt macroscopic phase transition ( $alpha > 0.62$ ) provide critical theoretical insights into sociotechnical transitions. Traditional policy evaluation frameworks predominantly assess intervention measures in isolation. However, our multiplex ABM demonstrates that in complex human-environment systems characterized by subsurface invisibility, policy interventions interact non-linearly. Without preemptive bias reduction (Policy A) and network rumor suppression (Policy D), increasing procedural transparency (Policy B) inadvertently amplifies public sensitivity to operational noise, demonstrating that transparency without foundational trust is counterproductive.

## 4.2. Alignment with Real-World Sociotechnical Cases

Our computational findings exhibit striking qualitative alignment with empirical observations across historical energy deployment conflicts:

(1) Illusory Avalanches: The phenomenon where ungrounded rumors ignite system-wide rejection cascades without physical plume interference mirrors public backlash in shale gas hydraulic fracturing [51] and visual/health nocebo effects in wind energy [32], where fear diffuses independently of physical harm.

(2) The Transparency Paradox: The failure of Policy B + E directly reflects empirical governance failures during European smart meter rollouts [30] and theoretical warnings by [77] and [46], proving that disclosing technical data to a deeply distrustful public heightens hypersensitivity rather than building reassurance.

## 4.3. Policy Implications and Implementation Feasibility

From a practical governance standpoint, relying on standalone economic compensation (Policy E) is structurally bounded; in high-trauma regimes ( $P(H0) > 0.40$ ), financial incentives are easily overwhelmed by cumulative protest utility. To steer ATES diffusion onto a sustainable trajectory, municipalities must deploy a fully integrated, multi-layered policy package. While simultaneous high-intensity implementation may impose administrative resource constraints, our model highlights that preemptive communication (Policy A) and digital twin firewalls (Policy D) serve as essential prerequisite investments that dramatically lower the required magnitude of financial compensation.

### 4.4. Methodological Limitations and Empirical Roadmap

As an "exploratory ABM", this study prioritizes uncovering systemic mechanisms and phase boundaries rather than site-specific quantitative forecasting. To translate these computational insights into real-world projects, future research should execute a two-stage empirical roadmap: conducting stated-preference choice conjoint surveys in prospective candidate cities to measure local loss-aversion coefficients and baseline prior distrust ($P(H_0)$), followed by calibrating these empirical metrics directly into the ABM parameter space.

## 5. Conclusion

From the analytical findings of this model, policymakers must recognize two fundamental insights:

1. Policy interventions exhibit diverse interaction dynamics—including synergistic combinations, offset combinations where one policy nullifies another, independent combinations, and complementary combinations that mitigate mutual shortcomings.
2. In sociotechnical contexts where human cognitive biases regarding invisible subsurface spaces interact with complex network dynamics, **an incomplete policy package will either lead to results identical to the baseline (a lock-in state of 0% adoption) or actively worsen public resistance, as exemplified by the monitoring paradox.**

To completely invalidate "illusory avalanches" and steer renewable technology diffusion onto a sustainable trajectory, deploying a "fully integrated, multi-layered policy package" that simultaneously covers cognitive (Policy A), institutional (Policies B and C), network (Policy D), and economic (Policy E) dimensions is paramount. Even when immediate full implementation is constrained by administrative resources, simulating pairwise and multi-layered policy package combinations prior to real-world deployment is strongly advised.

On the other hand, while the utility functions and Bayesian updating parameters in this model are directly grounded in stylized facts established in behavioral economics—such as loss aversion under Prospect Theory [80]—and the mathematical model parameters are structurally mapped onto frameworks of procedural justice, distributive justice, and institutional justice [15] [46], future research should advance an empirical validation roadmap. Specifically, choice-experiment choice conjoint surveys should be conducted across prospective candidate cities to empirically measure residents' loss aversion coefficients and baseline distrust levels, subsequently calibrating these empirical metrics as initial input parameters within this ABM.

## 6. Conclusion and Methodological Roadmap for Empirical Validation

Rather than functioning as a predictive forecasting tool, this study serves as an "exploratory ABM" (Epstein, 2006) explicitly designed to map non-linear phase boundaries, counter-intuitive trade-offs (such as the monitoring paradox), and policy synergy thresholds in complex sociotechnical systems. By mathematically unifying subsurface thermal advection-dispersion with social Bayesian belief updating, this research addresses a critical gap in energy social science: explaining how subterranean physical dynamics and surface digital communication networks co-evolve to shape renewable energy acceptance.

To translate these computational insights into site-specific policy execution, we propose a concrete, two-stage empirical validation roadmap:

- Empirical Calibration via Choice Conjoint Experiments: Stated-preference conjoint surveys will be administered to prospective host communities to empirically quantify local loss-aversion coefficients, baseline prior distrust $P(H_0)$, and exact monetary-to-utility conversion rates.
- In-situ ABM Parameterization: Empirical utility distributions derived from conjoint choice tasks will replace synthetic parameter priors, enabling municipal planners to simulate location-specific critical policy thresholds ($\alpha_{\text{critical}}$) before breaking ground.

In conclusion, navigating the deep decarbonization of urban space demands moving beyond isolated, single-dimensional policy fixes. Deploying a fully integrated, multi-layered policy package—simultaneously addressing cognitive biases, institutional guarantees, network rumor dynamics, and distributive justice—is paramount to dissolving social friction and securing the long-term societal acceptance of subsurface energy technologies.

## Appendix A. Sensitivity Analysis of Behavioral Thresholds and Weight Parameters

To verify whether the observed macroscopic phenomena—specifically the "monitoring paradox" in Policy B+E and the rapid phase transition under the Full Package (Policies A–E)—are robust against arbitrary parameter choices, we executed a sensitivity analysis across the parameter space of the protest threshold ($U_{\text{protest}} \in [8.0,22.0]$) and the cognitive weight ratio ($w_{\text{distrust}}/(w_{\text{distrust}} + w_{\text{social}}) \in [0.1,0.9]$). As shown in Figure A.1 (left), under the Policy B+E regime at policy strength $\alpha = 0.7$, adoption collapses to near zero across the majority of parameter combinations, confirming that the monitoring paradox is a structural artifact of Bayesian updating under high transparency rather than a parameter-specific outcome. Conversely, Figure A.1 (right) demonstrates that under the fully integrated policy package at $\alpha = 0.7$ ($> \alpha_{critical} = 0.62$), system adoption is uniformly sustained at nearly 100% across virtually the entire parameter grid. These findings confirm the structural robustness of our model's core theoretical conclusions.

**Fig.A1 Sensitivity Analysis**

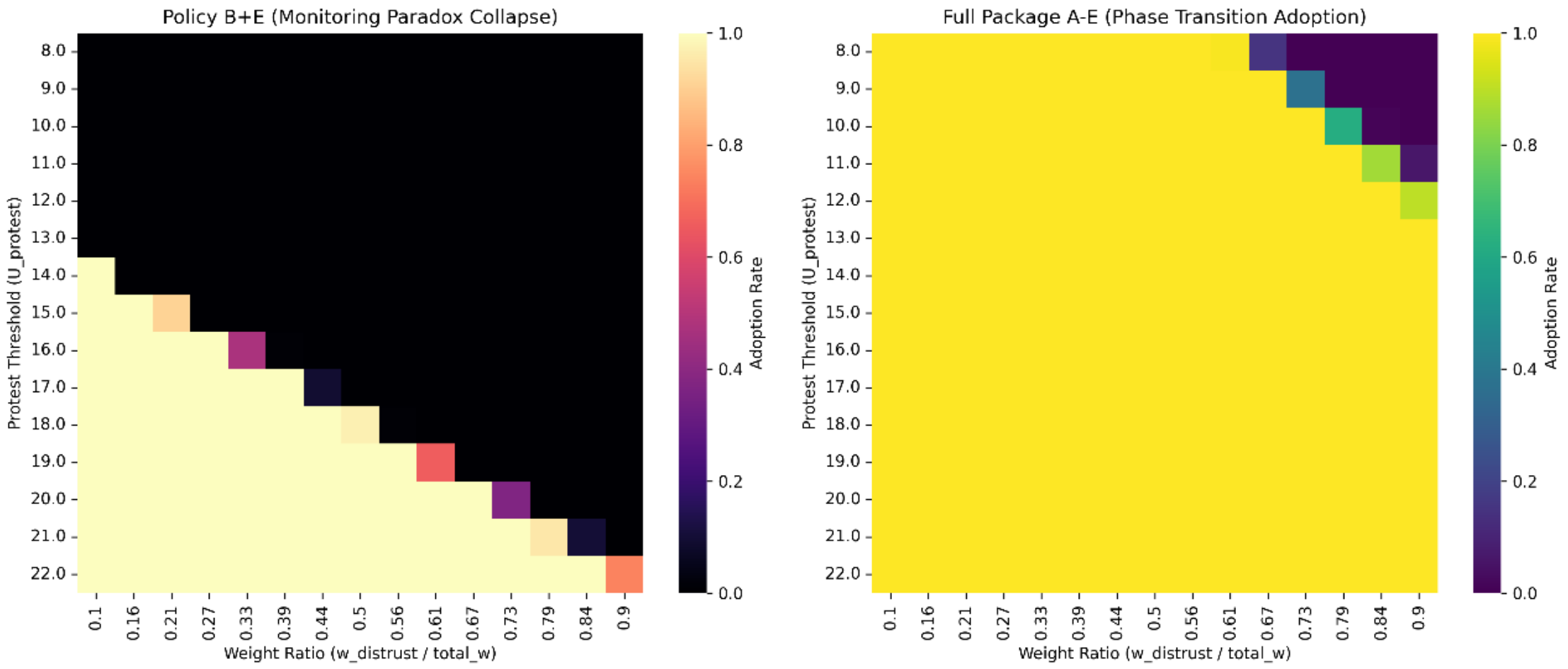